\documentstyle[pre,aps]{revtex}

\def\eps{\varepsilon}
\def\partt{\mbox{\boldmath $\partial$}}
\def\h{{\bf h}}

\def\k{{\bf k}}
\def\x{{\bf x}}
\def\r{{\bf r}}

\begin{document}
\draft

 \title{``TOY MODELS'' OF TURBULENT CONVECTION AND
THE HYPOTHESIS OF THE LOCAL ISOTROPY RESTORATION }

 \author{N. V. Antonov}
 \address{
 Department of Theoretical Physics, St~Petersburg University,
 Uljanovskaja 1, St~Petersburg, Petrodvorez, 198904 Russia}


 \maketitle

\vskip1cm
\hfill{\it Dedicated to the memory of Anatolij Georgievich Izergin}
\vskip1cm

\begin{abstract}
A brief review is given of recent results devoted to the effects of
large-scale anisotropy on the inertial-range statistics of the
passive scalar quantity $\theta(t,{\bf x})$, advected by the synthetic
turbulent velocity field with the covariance
$\propto\delta(t-t')|{\bf x}-{\bf x'}|^{\eps}$. Inertial-range anomalous
scaling behavior is established, and explicit asymptotic expressions for
the structure functions $ S_n (\r) \equiv \langle
[\theta(t,{\bf x}+\r)-\theta(t,{\bf x})]^{n} \rangle $
are obtained; they are represented by superpositions of power laws with
universal (independent of the anisotropy parameters) anomalous exponents,
calculated to the first order in $\eps$ in any space dimension.
The exponents are associated with tensor composite operators
built of the scalar gradients, and exhibit a kind of hierarchy related
to the degree of anisotropy: the less is the rank, the less is the
dimension and, consequently, the more important is the contribution
to the inertial-range behavior. The leading terms of the even (odd)
structure functions are given by the scalar (vector) operators.
The small-scale anisotropy reveals itself in odd correlation functions:
for the incompressible velocity field,
$S_{3}/S_{2}^{3/2}$ decreases going down towards to the depth of
the inertial range, while the higher-order odd ratios increase;
if the compressibility is strong enough, the skewness factor also
becomes increasing.

This paper is intended for the memorial issue of ``Zapiski Nauchnykh
Seminarov POMI'' dedicated to the memory of Anatolij Georgievich Izergin.

\end{abstract}


The investigation of intermittency and anomalous scaling in fully
developed turbulence remains essentially an open theoretical problem.
Both the natural and numerical experiments suggest that the deviation
from the predictions of the classical Kolmogorov--Obukhov theory
\cite{Monin,Legacy} is even more strongly pronounced for a passively
advected scalar field than for the velocity field itself; see, e.g.,
\cite{An,synth} and literature cited therein. At the same time,
the problem of passive advection appears to be easier tractable
theoretically: even simplified models describing the advection by
 ``synthetic'' velocity fields with given Gaussian statistics
reproduce many of the anomalous features of genuine turbulent
heat or mass transport observed in experiments. Hence the term
``toy models.'' Therefore, the
problem of passive scalar advection, being of practical
importance in itself, may also be viewed as a starting point
in studying anomalous scaling in the turbulence on the whole.

The most remarkable progress has been achieved for the so-called
rapid-change model, which dates back to Obukhov and Batchelor and was
explicitly introduced by Kraichnan \cite{Kraich1}: for the first time, the
anomalous exponents have been calculated on the basis of a microscopic
model and within regular perturbation expansions; see, e.g., [6--10]
and references therein.

Within the ``zero-mode approach,''  developed in [6--9], nontrivial anomalous
exponents are related to the zero modes (unforced solutions)
of the closed exact equations satisfied by the equal-time
correlations. Within the approach based on the field theoretic
renormalization group (RG) and operator product expansion (OPE),
the anomalous scaling emerges as a consequence of the existence in the
model of composite operators with {\it negative} critical dimensions,
which determine the anomalous exponents [10--13].

An important question recently addressed is the effects of large-scale
anisotropy on the inertial-range statistics of passively ad\-vec\-ted fields
[8,9,12--18] and the velocity field itself \cite{Arad1}.

According to the classical Kolmogorov--Obukhov theory \cite{Monin,Legacy},
the anisotropy introduced at large scales by the forcing (boundary
conditions, geometry of an obstacle {\it etc}) dies out when the energy is
transferred down to smaller scales owing to the cascade
mechanism. A number of recent works confirms this picture
for the {\it even} correlation functions, thus giving some quantitative
support to the aforementioned hypothesis on the restored local isotropy
restoration of the inertial-range turbulence for the velocity and passive
fields [12--19]. More precisely, the exponents describing the inertial-range
scaling exhibit universality and hi\-e\-ra\-r\-chy related to the degree of
anisotropy, and the leading contribution
to an even function is given by the exponent from the isotropic shell
[12,15--19]. Nevertheless, the anisotropy survives
in the inertial range and reveals itself in {\it odd} correlation functions,
in disagreement with what was expected on the basis of the cascade ideas.
The so-called skewness factor decreases down the scales much slower than
expected \cite{An,synth,Pumir,Siggia}, while the higher-order odd
dimensionless ratios (hyperskewness {\it etc}) increase, thus signalling of
persistent small-scale anisotropy \cite{RG3,CLMV99,Lanotte2}. The effect
seems rather universal, being observed for the scalar \cite{RG3} and vector
\cite{Lanotte2} fields, advected by the Gaussian rapid-change
velocity, and for the scalar advected by the two-dimensional Navier-Stokes
(NS) velocity field \cite{CLMV99}.

Below we briefly review the most recent results obtained on the example of
the rapid-change model using the RG and OPE techniques; more detailed
discussion can be found in the original papers [10--13].
The detailed description of the RG approach to the models of fully
developed turbulence is given in the book \cite{turbo} and review paper
\cite{UFN}. For simplicity, we restrict ourselves with the rapid-change
models; the case of a finite-correlated velocity field is considered
\cite{RG3,RG4}.

The advection of a passive scalar field in the presence of an
imposed linear gradient is described by the equation
\begin{equation}
\partial _t\theta+ ({\bf v}\partt) \theta
=\nu _0\partial^{2} \theta-(\h{\bf v}) .
\label{1}
\end{equation}
Here $\theta(x)\equiv \theta(t,{\bf x})$ is the random (fluctuation)
part of the total scalar field $\Theta(x)=\theta(x)+(\h{\bf x})$,
$\h$ is a constant vector that determines distinguished direction,
$\nu _0$ is the molecular diffusivity coefficient,
$\partial _t \equiv \partial /\partial t$,
$\partial _i \equiv \partial /\partial x_{i}$, and
$\partial^{2}\equiv\partial _i\partial _i$
is the Laplace operator.

The velocity field ${\bf v}(x)=\{v_i(x)\}$
obeys a Gaussian distribution with zero mean and correlator
\begin{equation}
\langle v_{i}(x) v_{j}(x')\rangle = \delta(t-t')\, K_{ij}(\r)
\label{temporal}
\end{equation}
with
\begin{equation}
K_{ij} (\r) = D_{0}\,
\int \frac{d{\bf k}}{(2\pi)^d} \,
\frac{P_{ij}({\bf k})} {k^{d+\eps}} \,
\exp [{\rm i}{\bf k}\cdot\r] ,
\quad \r\equiv {\x}-{\x'},
\label{spatial}
\end{equation}
where $P_{ij}({\bf k})=\delta_{ij}-k_i k_\j/
k^2$ is the transverse projector, $\k$ is the momentum, $k\equiv|{\bf k}|$,
$D_{0}>0$ is an amplitude factor, and $0<\eps<2$
is a free parameter. The IR regularization is provided by
the cut-off in the integral (\ref{spatial}) from below at
$k\simeq m$, where $m\equiv 1/L$ is the reciprocal of the
integral turbulence scale; the precise form of the cut-off
is not essential. For $0<\eps<2$, the difference
\begin{equation}
S_{ij}(\r) \equiv K_{ij} ({\bf 0}) - K_{ij} (\r)
\label{difference}
\end{equation}
has a finite limit for $m\to0$:
\begin{equation}
S_{ij}(\r)= D r^{\eps}\left [\left (d+\eps-1\right)
\delta_{ij} - \eps \frac{r_{i}r_{j}}{r^2}  \right],
\label{eddydiff}
\end{equation}
with
\[ D = \frac{- D_{0}\, \Gamma(-\eps/2)} {(4\pi)^{d/2}2^{\eps}(d+\eps)
\Gamma(d/2+\eps/2)}, \]
where $\Gamma(\cdots)$ is the Euler gamma function
(note that $D>0$). It follows from Eq. (\ref{eddydiff})
that $\eps$ can be viewed as a kind of H\"{o}lder exponent,
which measures the {\em roughness\/} of the velocity field.
In the RG approach, the exponent $\eps$ plays the same
role as the parameter $\varepsilon=4-d$ does in the RG theory
of critical phenomena \cite{Zinn,book3}. The relations
\begin{equation}
g_{0} \equiv D_{0}/\nu_0 \equiv \Lambda^{\eps}
\label{g0}
\end{equation}
define the coupling constant $g_{0}$ (i.e., the expansion parameter
in the ordinary perturbation theory) and the characteristic
ultraviolet (UV) momentum scale $\Lambda$.

The term $(\h{\bf v})$ in Eq. (\ref{1}) acts as a driving force; it
maintains the steady state and, simultaneously, introduces the large-scale
anisotropy. In the original (isotropic) formulation of the model
this term is replaced with an artificial Gaussial random force $f(x)$
with the covariance
\begin{equation}
\langle  f(x)  f(x')\rangle = \delta(t-t')\, C(mr), \quad
r\equiv|{\bf x}-{\bf x'}|,
\label{2}
\end{equation}
where $C(mr)$ is a function finite as $mr=0$.

The issue of interest is, in particular, the behavior of the
equal-time structure functions
\begin{equation}
S_{n}(\r)\equiv\langle[\theta(t,{\bf x})-\theta(t,{\bf x'})]^{n}\rangle
\label{struc}
\end{equation}
in the inertial range, specified by the inequalities
$ \Lambda >>1/r >>m $. The dimensionality considerations give
\begin{equation}
S_{n}(\r)=  \nu_0^{-n/2}\, r^{n}\, R_{n} (\Lambda r, mr, z),
\label{strucdim}
\end{equation}
where $z$ is the cosine between the directions of $\r$ and $\h$ and
$R_{n}$ are some functions of dimensionless parameters. In
principle, they can be calculated within the ordinary perturbation
theory (i.e., as series in $g_{0}$), but this is not useful for
studying inertial-range behavior: the coefficients
are singular in the limits $\Lambda r \to\infty$  and/or
$mr\to0$, which compensate the smallness of $g_{0}$
(assumed in ordinary perturbation theory),
and in order to find correct IR behavior we have to sum the entire
series. The desired summation can be accomplished using
the RG and OPE; see Refs. \cite{RG,RG1,RG3}.

The RG analysis consists of two main stages. On the
first stage, the multiplicative renormalizability of the model is
demonstrated and the differential RG equations for its correlation
functions are obtained. The asymptotic behavior of the functions
like (\ref{struc}) for $\Lambda r>>1$ and any fixed $mr$ is given
by IR stable fixed points of the RG equations and has the form
\begin{equation}
S_{n}(r)= D_0^{-n/2}\, r^{n(1-\eps/2)}\, R_{n} (mr,z), \quad \Lambda r>>1,
\label{strucdim2}
\end{equation}
with certain, as yet unknown, scaling functions $R_{n} (mr,z)$.

On the second stage, the small $mr$ behavior of the functions
$ R_{n} $ is studied within the general representation
(\ref{strucdim2}) using the OPE. It shows that, in the limit $mr\to0$,
the functions $ R_{n}$ have the asymptotic forms
\begin{equation}
 R_{n} (mr,z) = \sum_{F} C_{F}(mr,z)\, (mr)^{\Delta_{F}},
\label{ope}
\end{equation}
where $C_{F}$ are coefficients regular in $mr$. In general,
the summation is implied over all possible renormalized composite
operators $F$ allowed by the symmetry; $\Delta_{F}$ being their
critical dimensions.

The peculiarity of the models describing turbulence is the existence
of the so-called ``dangerous'' composite operators with {\it negative}
critical dimensions \cite{RG,RG1,RG3,turbo,UFN}. Their contributions
into the OPE give rise to a singular behavior of the scaling functions for
$mr\to0$, and the leading term is given by the operator with minimal
$\Delta_{F}$. In the isotropic case only scalar operators have
nonvanishing mean values and contribute to the right hand side of Eq. (\ref{ope}).
The odd functions $S_{2k+1}$ vanish, while the leading contributions
to $S_{2k}$ are determined by scalar gradients
$F_{k} = (\partial_{i}\theta\partial_{i}\theta)^{k}$ and have the form
\begin{equation}
S_{2k}(r) \simeq D_{0}^{-k}\, r^{k(2-\eps)}\, (mr)^{\Delta_{k}},
\label{HZ1}
\end{equation}
where the critical dimensions $\Delta_{k}$ of the operators $F_{k}$
are given by
\begin{equation}
\Delta_{k}=-2k(k-1)\eps/(d+2)+O(\eps^{2})=-2k(k-1)\eps/d+O(1/d^{2}).
\label{exps}
\end{equation}
The expression
(\ref{HZ1}) agrees with the results obtained earlier in \cite{Falk1,GK}
using the zero-mode techniques; the $O(\eps^{2})$ contribution to
$\Delta_{k}$ is reported in \cite{RG}.
Note that the operators $F_{l}$ with $l\ge k$ (whose contributions would be
more important) do not appear in the right hand side of the OPE for
$S_{2k}$; this is a consequence of the linearity of the original equation
(\ref{1}) in $\theta(x)$.

In the anisotropic case, an important role is played by the
irreducible tensor operators of the form
\begin{equation}
F[n,p]\equiv \partial_{i_{1}}\theta\cdots\partial_{i_{p}}\theta\,
(\partial_{i}\theta\partial_{i}\theta)^{l} +\cdots,
\label{Fnp}
\end{equation}
where $p$ is the number of the free vector indices and $n=p+2l$
is the total number of the fields $\theta$ entering into the operator;
the vector indices of the symbol $F[n,p]$ are omitted.
The dots stand for the subtractions (needed
for $p\ge2$) which ensure that the resulting expression is traceless
with respect to any contraction, for example,
$F[2,2]= \partial_{i} \theta \partial_{j} \theta - \delta_{ij}
F[2,0]/d$.

For $\h\ne0$, the operators (\ref{Fnp}) acquire nonvanishing mean values,
built of the vectors $\h$ and delta symbols,
and give nontrivial contributions to the OPE for $S_{n}$;
in particular, this gives rise to nonvanishing odd functions.

The critical dimensions associated with the operators (\ref{Fnp})
have the form \cite{RG3} (see also Refs. \cite{SOd,Wiese}):
\begin{equation}
\Delta[n,p] = \frac{2n(n-1) - (d+1)(n-p) (d+n+p-2)}{2(d-1)(d+2)}\,
\eps +O( \eps^{2}).
\label{Dnp}
\end{equation}

The straightforward analysis of the explicit one-loop
expression (\ref{Dnp}) shows that for any fixed $n$,
the dimension $\Delta[n,p]$
decreases monotonically with $p$ and reaches its minimum for
the minimal possible value of $p=p_{n}$, i.e., $p=0$ if $n$ is
even and $p=1$ if $n$ is odd:
\begin{equation}
\Delta[n,p] > \Delta[n,p']\quad  {\rm if} \quad  p>p'.
\label{hier2}
\end{equation}
Furthermore, this minimal value $\Delta[n,p_{n}]$ is
negative and decreases monotonically as $n$ increases:
\begin{equation}
0>\Delta[2k,0]>\Delta[2k+1,1]>\Delta[2k+2,0].
\label{hier3}
\end{equation}
Finally, we note that for any fixed $p$, the dimension (\ref{Dnp})
decreases monotonically as $n$ increases:
\begin{equation}
\Delta[n,p] > \Delta[n',p]\qquad  {\rm if} \quad  n<n'.
\label{hier4}
\end{equation}

The mean value of the operator with the dimension $\Delta[k,p]$
is a traceless (for $p\ge2$) tensor built of the vector $\h$ and
Kronecker delta symbols; the contraction with the vector indices
of the corresponding coefficient $C_{F} $ in Eq.
(\ref{ope}) gives rise to $p$-th order Legendre polynomial
$P_{p}(z)$, so that the expansion (\ref{ope}) is
analogous to the decomposition in irreducible representations of
the rotation group, used in Refs. [15--19].

The inequalities (\ref{hier2}), (\ref{hier3}) show that the
contributions of the tensor operators (\ref{Fnp}) into the
asymptotic expression (\ref{ope}) exhibit a kind of hierarchy:
the less is the rank, the more important is the contribution.

The leading term of the expression (\ref{ope}) for the even (odd)
function $S_n$ is determined by the scalar (vector) composite
operator consisting of $n$ factors $\partial\theta$
and has the form
\begin{equation}
S_{n}(r) \propto (hr)^n\, (mr)^{\Delta[n,p_{n}]} .
\label{struc3}
\end{equation}

Let us now explore the consequences of our results for the issue of the
local isotropy restoration in this problem.

As a measure of small-scale anisotropy one usually uses the dimensionless
ratios of the structure functions:
\begin{equation}
R_{n}(\r) \equiv {S_{n}(\r)}/{S_{2}^{n/2}(\r)} .
\label{pair-n}
\end{equation}
From Eq. (\ref{struc3}) it then follows that in inertial range we have:
\begin{equation}
R_{2k+1} \propto  (mr) ^{\Delta[{2k+1,1}]-(2k+1)\Delta[{2,0}]/2}\,,
\label{pair-odd}
\end{equation}
\begin{equation}
R_{2k} \propto (mr) ^{\Delta[2k,0]-k\Delta[{2,0}]} .
\label{pair-even}
\end{equation}
The dependence
on the P\'eclet number, $Pe \equiv (\Lambda/m)^{\eps}$  can be estimated
by replacing $r$ with $1/\Lambda$; see Ref.~\cite{Pumir}.
Using explicit $O(\eps)$ expressions for $\Delta[{n,p}]$ we then obtain:
\begin{equation}
R_{2k+1} \propto Pe^{-(d+2-4k^{2})/[2(d+2)]},
\label{pair-11}
\end{equation}
\begin{equation}
R_{2k} \propto Pe^{2 k(k-1)/(d+2)}.
\label{pair-12}
\end{equation}

Since the leading terms of the even functions (\ref{struc3}) are
determined by the exponents of the isotropic shell (i.e., those related
to the scalar composite operators), the inertial-range behavior of
the even ratios (\ref{pair-even}), (\ref{pair-12}) is the same as in
the isotropic model.
This gives a quantitative support to the universality of
{\it leading} anomalous exponents for the {\it even} functions
with respect to different classes of forcing.
On the other hand, the odd quantities (\ref{pair-odd}), (\ref{pair-11})
appear sensitive to the anisotropy: $R_{3}$ in (\ref{pair-11})
slowly decreases for $Pe\to\infty$, while ratios $R_{2k+1}$ with $k\ge 2$
{\it increase} with $Pe$, thus signalling of the persistent small-scale
anisotropy.

The picture outlined above seems rather general. Indeed, it is compatible
with that recently established for the NS turbulence \cite{Arad1},
the scalar field passively advected by a NS velocity in the two-dimensional
inverse energy cascade \cite{CLMV99}, and for the vector
(magnetic) field advected by the velocity of the type (\ref{spatial}),
see Refs. \cite{Lanotte,Lanotte2,ABP}.

To conclude with, let us briefly discuss the effects of compressibility
onto the small-scale anisotropy persistence. The transversal projector in
the correlator (\ref{spatial}) is then replaced by the combination
$$P_{ij}({\bf k})+\alpha Q_{ij}({\bf k}),$$
where $Q_{ij}({\bf k}) = k_i k_j / k^2$ is the longitudinal projector
and $\alpha>0$ is a free parameter.
The model (\ref{1}) then corresponds to the advection of a tracer
(entropy, temperature, concentration of a pollutant), while the
nonlinearity $\theta' \partt ({\bf v}\theta)$ describes advection
of a density.

Equation (\ref{Dnp}) for a tracer becomes  \cite{RG4}
\begin{equation}
\Delta[n,p] = \frac{2n(n-1)(1-\alpha)-(n-p) (n+p+d-2)
(d+1+\alpha)}{2(d+2)(d-1+\alpha)} \,\eps ,
\label{DnpA}
\end{equation}
the $O(\eps^{2})$ correction is calculated in \cite{Juha}
for all $n$, $p$, $d$ and $\alpha$.

Although the hierarchy relations (\ref{hier2})--(\ref{hier4})
remain valid for all values of $\alpha>0$
(in particular, $\partial \Delta[n,p] / \partial p>0$), the corrections
become closer to leading terms as $\alpha$ increases:
$$\partial^{2} \Delta[n,p] / \partial p \partial\alpha<0.$$
Furthermore, the expressions (\ref{pair-11})
take on the form
\begin{equation}
R_{2k+1} \propto (mr)^{\eps \,
[(d-1+\alpha)(d+2-4k^{2})-8\alpha\, k^{2}] /2(d+2)(d-1+\alpha)}.
\label{pair-11A}
\end{equation}
It follows from (\ref{pair-11A}) that $R_{3}$ diverges for
$mr\to0$ provided $\alpha$ is large enough [namely,
$\alpha>(d-1)(d+2)/(10-d)+O(\eps)$], while the divergence of the
higher ratios $R_{2k+1}$ enhances as $\alpha$ increases.

This means that compressibility enhances the penetration of the
large-scale anisotropy towards the depth of the inertial range.
This fact also seems universal, being observed in the model of
the passively advected magnetic field \cite{Paolo}.

Some remarks are now in order.

The critical dimensions of all composite operators (\ref{Dnp})
are independent of the forcing: they remain unchanged, when the
isotropic stirring force (\ref{2}) is replaced in Eq. (\ref{1})
with the anisotropic term $(\h{\bf v})$. The difference is that for the
former case, only scalar operators contribute to the representations
like (\ref{ope}), while for the latter case
the irreducible tensor operators acquire nonzero mean values and
also contribute to  Eq. (\ref{ope}).
This is easily understood in the language of the zero-mode
approach: the noise $f$ and the term $(\h{\bf v})$ do not affect
the differential operators in the equations satisfied by the equal-time
correlations functions; the zero modes (homogeneous solutions)
coincide in the two cases, but in the latter case the modes with
nontrivial angular dependence should be taken into account.
In the language of the RG (which is also applicable to the case of a
finite-correlated velocity, see Refs. \cite{RG3,RG4}) this is explained
as follows: the stirring force or field $\h$ do not enter into the
diagrams that determine the renormalization of the operators
(\ref{Dnp}).
However, the anomalous exponents of the passive scalar become nonuniversal
and acquire the dependence on the anisotropy parameters if the velocity
field is taken to be strongly anisotropic also at small scales \cite{last}.

The picture outlined above and in Refs. \cite{Lanotte,Lanotte2} for
passively advected fields (a superposition of power laws with
universal exponents and nonuniversal amplitudes) seems rather
general, being compatible with that established recently in the
field of NS turbulence, on the basis of numerical
simulations of channel flows and experiments in the atmospheric
surface layer, see Refs. \cite{Arad1} and references therein.
In those papers, the velocity structure functions were decomposed
into the irreducible representations
of the rotation group. It was shown that in each sector of the
decomposition, scaling behaviour can be found with apparently universal
exponents. The amplitudes of the various contributions are nonuniversal,
through the dependence on the position in the flow, the local degree
of anisotropy and inhomogeneity, and so on \cite{Arad1}.

This is rather surprising because the equations for the correlation
functions in such cases are neither closed nor isotropic and homogeneous.
Although the hierarchy similar to Eq. (\ref{hier2}) is demonstrated by
the critical dimensions of certain tensor operators in the stirred NS
turbulence, see Ref. \cite{Triple} and Sec. 2.3 of \cite{turbo},
the relationship between them and the anomalous exponents is not
obvious there. One can thus speculate that the anomalous scaling for
the genuine turbulence can also appear a linear phenomenon.
It is worth recalling here that the so-called ``additive fusion rules,''
hypothesized for the NS turbulence in a number of papers,
Refs. \cite{Eyink},
and characteristic of the models with multifractal behavior
(see Ref. \cite{DL}), arise naturally in the context of the rapid-change
models owing to their {\it linearity}. The existing results for the
Burgers turbulence can also be interpreted
naturally as a consequence of similar fusion rules,
where only finite number of dangerous operators contributes to each
structure function, see Ref. \cite{Burg1}.

Of course, one should not insist too much on this bold assumption.

To conclude, let us compare briefly the situation for the
passively advected fields with the case of weak acoustic turbulence,
where the spectra can be obtained as solutions of the linear kinetic
equations (see Refs.~\cite{waves1,waves2}). For weakly dispersive waves
(e.g., with the dispersion law $\omega(k)\propto k^{1+\delta}$ with
$\delta<<1$), the anisotropy introduced by the large-scale forcing
enhances going down towards to the depth of the inertial range
\cite{waves1}. The hierarchy of the exponents related to the Legendre
decomposition is opposite to that established below and in
Refs.~\cite{RG3,Lanotte,Arad1}: anisotropic corrections decrease
slower for larger $p$'s \cite{waves1}. On the contrary, for the
nondispersive waves ($\delta=0$) the hierarchy of the exponents is
similar to that in our case, the anisotropic corrections decay
faster and faster with $p$ and the spectrum  tends to become isotropic
at small scales \cite{waves2}. To the best of our knowledge, no
information is available for the higher-order correlation functions
for such models. One can thus conclude that turbulent systems can exhibit
essentially different types of behavior with respect to the small-scale
isotropy restoration.

\vskip0.5cm
\noindent {\it Acknowledgments.}
I am thankful to L.\,Ts.\,Adzhemyan, M.\,Hnati\v{c}, J.\,Honkonen,
M.\,L\"{a}ssig, A.\,Mazzino, P.\,Muratore-Ginanneschi,
M.\,Yu.\,Nalimov and A.\,N.\,Vasil'ev for discussions.
The work was supported in part by the Russian Foundation for Fundamental
Research (Grant No. 99-02-16783) and the Grant Center for Natural Sciences
of the Russian State Committee for Higher Education
(Grant No. 97-0-14.1-30). A part of this work was performed
during my stay at the Department of Physics of the University of
Helsinki, financed by the Academy of Finland.

\end{document}